\documentclass[12pt]{article}
\usepackage{epsf,amssymb}
\usepackage{slashed}
\newcommand{\ft}[2]{{\textstyle\frac{#1}{#2}}}

\def\rmi{{\rm i}}
\def\rmd{{\rm d}}
\def\unity{{\mathchoice {\mathrm{1\mskip-4mu l}} {\mathrm{ 1\mskip-4mu l}}
{\mathrm{ 1\mskip-4.5mu l}} {\mathrm{ 1\mskip-5mu l}}}}
\newcommand{\SU}{\mathop{\rm SU}}
\newcommand{\SO}{\mathop{\rm SO}}
\newcommand{\U}{\mathop{\rm {}U}}
\newcommand{\USp}{\mathop{\rm {}USp}}

\newcommand{\Sp}{\mathop{\rm {}Sp}}

\csname @addtoreset\endcsname{equation}{section}

\begin{document}

\begin{titlepage}
\begin{flushright}
KUL-TF-01/20\\
hep-th/0110263
\end{flushright}
\vspace{.5cm}
\begin{center}
\baselineskip=16pt
{\LARGE    Special geometries, from real to quaternionic$^\dagger$  
}\\
\vfill
{\large Antoine Van Proeyen  
  } \\
\vfill
{\small Instituut voor Theoretische Fysica, Katholieke Universiteit
Leuven\\ Celestijnenlaan 200D, B-30001 Leuven}
\end{center}
\vfill
\begin{center}
{\bf Abstract}
\end{center}
{\small Special geometry is most known from 4-dimensional $N=2$
supergravity, though it contains also quaternionic and real geometries. In
this review, we first repeat the connections between the various special
geometries. Then the constructions are given starting from the
superconformal approach. Without going in detail, we give the main
underlying principles. We devote special attention to the quaternionic
manifolds, introducing the notion of hypercomplex geometry, being
manifolds close to hyperk\"{a}hler manifolds but without a metric. These are
related to supersymmetry models without an action.}
 \vfill \hrule width 3.cm {\footnotesize \noindent $^\dagger$ Contribution to the
 proceedings of the `Workshop on special geometric structures in string
theory', Bonn, 8-11/9/2001.}
\end{titlepage}
\tableofcontents{}
\section{Introduction}
Special geometry is related to the supersymmetric theories with 8 real
supercharges. In 4 dimensions this corresponds to $N=2$, and special
geometry was first found in this form~\cite{deWit:1984rz,deWit:1984pk}.
The first motivation to study $N=2$ supergravity was given
in~\cite{deWit:1980ug} as `Extended supergravity theories offer prospects
for unified theories of gravitation with other fundamental interactions.
Although much is know for $N=1$ supergravity, not many complete results
exist for higher-$N$ theories'. Thus the first motivation was just to find
more general theories than those that already existed at the time. The
corresponding geometry in rigid supersymmetry was investigated
in~\cite{Sierra:1983cc,Gates:1984py}. A new boost came with the advent of
Calabi--Yau compactifications of string theory. It was found that special
geometry occurs in the moduli space of these
manifolds~\cite{Seiberg:1988pf,Cecotti:1989qn,Ferrara:1989vp,Candelas:1991pi,
Candelas:1991qd,Strominger:1990pd}. E.g. in this period, a
coordinate-independent formulation of special geometry was
developed~\cite{Castellani:1990zd,D'Auria:1991fj}. In the research on
black holes a few years later, the $N=2$ theories were a popular tool,
and the special geometry was particularly
useful~\cite{Kallosh:1992ii,Ferrara:1995ih}. The research on dualities
started with the Seiberg--Witten papers, which were based on the use of
(rigid) special geometry~\cite{Seiberg:1994rs,Seiberg:1994aj}. The AdS/CFT
correspondence~\cite{Aharony:1999ti} gave new applications of special
geometries, and with brane-world scenarios as
in~\cite{Randall:1999ee,Randall:1999vf}, also the 5-dimensional variants
(`very special', see below) of special geometry got a lot of attention.
The relevant actions were first found in~\cite{Gunaydin:1984bi} and
connected to special geometry in~\cite{brokensi}. A full account of the
present knowledge was given in~\cite{AnnaGianguido}. The mathematicians
got interested in special geometry due to its relation with quaternionic
geometry~\cite{Cecotti:1989qn}, which lead to new results on the
classification of homogeneous quaternionic
spaces~\cite{deWit:1992nm,Cortes1996}.

Why are the theories with 8 supersymmetries so interesting? The maximal
supergravities\footnote{The restriction is due to interacting field theory
descriptions, which e.g. in 4 dimensions does not allow fields with spin
larger than~2.} contain 32 supersymmetries. These are the $N=8$ theories
in 4 dimensions, and exist in spaces of Lorentzian signature with at most
11 dimensions, i.e. (10,1) space--time dimensions. If one allows more time
directions, 32 supersymmetries is possible in 12 dimensions with (10,2)
or (6,6) signature. However, these theories allow no matter multiplets
(multiplets with fields up to spin~1). For the geometry, determined by
the kinetic terms of the scalars, this means that the manifold is fixed
once the dimension is given. For all theories with 32 supersymmetries this
is a symmetric space.

Matter multiplets are possible if one limits the number of
supersymmetries to 16 (thus $N=4$ in 4 dimensions). These theories exist
up to 10 dimensions with Lorentzian signature. In this case, the geometry
is fixed to a particular coset geometry once one gives the number of
matter multiplets that are coupled to supergravity.

The situation becomes more interesting if the number of supersymmetries
is~8. Now there are functions, which can be varied continuously, that
determine the geometry. This makes the geometries much more interesting.
Of course, if one further restricts to 4 supersymmetries, more geometries
would be possible. In 4 dimensions, e.g., general K\"{a}hler manifolds
appear. For 8 supersymmetries, these are restricted to 'special K\"{a}hler
manifolds', determined by a holomorphic prepotential. However, this
restriction makes the class of manifolds very interesting and manageable.
The holomorphicity is a useful ingredient, and was e.g. essential to
allow the solution of the theory in the Seiberg--Witten
model~\cite{Seiberg:1994rs,Seiberg:1994aj}. The theories with 8
supersymmetries are thus the maximally supersymmetric that are not
completely determined by the number of fields in the model, but allow
arbitrary functions in their definition, i.e. continuous deformations of
the metric of the manifolds.

In the following section, we will give an overview of the matter
multiplets with 8 supersymmetries. Then, in
section~\ref{ss:familyspecial} we will present the family of special
geometries. Section~\ref{ss:scForExtra} will show why superconformal
methods are useful when one wants to understand the extra symmetries
(related to duality) that appear in special geometry. This sets the stage
for the main section (section~\ref{ss:construction3}) where we will
explain the construction of the models using 3 steps: introduction of the
multiplets in the superconformal framework, construction of the action,
and gauge fixing. We will devote special emphasis to hypermultiplets and
quaternionic geometry. The presentation here is new, and introduces
models without an action, related to hypercomplex geometry. We give some
conclusions in section~\ref{ss:conclusions}.


\section{Matter multiplets with 8 supersymmetries}
The maximal spacetime for 8 supersymmetries is 8-dimensional, with
signature (8,0) or (4,4). We restrict ourselves to Lorentzian signature,
and as such the maximal dimension is~6. An overview of the matter
multiplets is given in table~\ref{tbl:matter8susy}.
\begin{table}[ht]
  \caption{\it Matter multiplets with 8 supersymmetries}\label{tbl:matter8susy}
\begin{center}
\begin{tabular}{|c|c|c|c|}
\hline
$D=6$  & $D=5$ & $D=4$ & $D=3$ \\
\hline
tensor multiplet  &  &  &  \\
$\mathbb{R}$  & vector/tensor  & vector multiplet&  \\
\cline{1-1}
vector multiplet  & $\mathbb{R}$ & $\mathbb{C}$ & hypermultiplet \\
\small \textit{no scalar}  &  &  & $\mathbb{Q}$ \\  \cline{1-3}
hypermultiplet  & hypermultiplet & hypermultiplet &  \\
$\mathbb{Q}$  &$\mathbb{Q}$  & $\mathbb{Q}$ &  \\
\hline
\end{tabular}
\end{center}
\end{table}
We now discuss the geometries defined by the kinetic action of the scalar
fields in these multiplets.

In 6 dimensions there are tensor multiplets, vector multiplets and
hypermultiplets. The former have a 2-form gauge field with self-dual
field strength. They have been studied first in~\cite{Romans:1986er}. The
scalars of $n$ tensor multiplets coupled to supergravity describe a coset
space SO$(1,n)/$SO$(n)$. They can be coupled to vector multiplets. These
couplings are governed by constants $c_{IMN}$, where $I$ labels the
tensor multiplets, and $M$ and $N$ the vector multiplets, but these
couplings do not influence the geometry of the scalar manifolds, as vector
multiplets in 6 dimensions do not contain scalars. A full account of
these couplings is given in~\cite{Riccioni:2001bg}. The hypermultiplets
determine a separate sector of the manifold.

Dimensional reduction deforms a multiplet in table~\ref{tbl:matter8susy}
to the one to its right. Vector and tensor multiplets appear in one box in
5 dimensions, as Abelian vectors are dual to antisymmetric tensors in
$D=5$. As long as we only discuss the geometry, we do not have to make a
distinction. When one considers gauged symmetry groups then the vector
and tensor multiplets play a different role~\cite{GunZag}, but for the
geometry they can be considered as equivalent. They reduce to vector
multiplets in 4 dimensions, as illustrated in
table~\ref{tbl:reduceMatter}.
\begin{table}[ht]
  \caption{\it Reduction of matter multiplets between 5,4 and 3 dimensions.}
  \label{tbl:reduceMatter}
\begin{center}
  \begin{tabular}{|l|ccccccc|}
\hline
  & $D=5$  &  & $D=4$ &  & \multicolumn{3}{c|}{$D=3$}  \\
  &        &  &       &  &  &  & after  \\
  &        &  &       &  &  &  &duality \\
\hline
spin $1/2$  & 2 & $\rightarrow$ & 2 & $\rightarrow$ & 4 & $\rightarrow$& 4  \\[3mm]
spin 1    & 1 & $\rightarrow$ & 1 & $\rightarrow$ & 1 & &  \\[-3mm]
          &   & $\searrow$    &   & $\searrow$    & &$\searrow$&  \\
spin 0    & 1 & $\rightarrow$ & 2 & $\rightarrow$ & 3 & $\rightarrow$ & 4
\\[4mm]
     &$\mathbb{R}$ &     & $\mathbb{C}$ &     &  &  & $\mathbb{Q}$ \\
\hline
\end{tabular}
\end{center}
\end{table}
These have complex scalars, which can be seen as the 5th and 6th
components of the vectors of 6 dimensions. In 3 dimensions, Abelian
vectors are dual to scalars. This leads to the last column of
table~\ref{tbl:reduceMatter}. Thus, in 3 dimensions all the multiplets
are equivalent to hypermultiplets, i.e. the multiplets with only spin-0
and spin-1/2 fields.

The hypermultiplets in all the dimensions look alike. Indeed, the scalar
sector does not depend on the spacetime dimension. There are in any case
4 scalars for each multiplet. As we will show below, they appear in a
quaternionic structure. The spin-1/2 fields are differently organised in
3,4 and 5 dimensions, but that is not visible in the geometry.

Therefore, in the following we will concentrate on the upper line of
table~\ref{tbl:matter8susy}, starting from the vector/tensor multiplets
in 5 dimensions with real scalars, the vector multiplets in 4 dimensions
with complex scalars, and the hypermultiplets in 3 dimensions (and which
describe thus the geometry as well for 4,5 and 6 dimensions) with
quaternionic scalars. We will leave the rather obvious geometry of the
tensor multiplets in 6 dimensions aside.

When one considers the supergravity theory, there is an extra
contribution from the reduction of the supergravity multiplet. The
supergravity multiplets in 5 dimensions (as in 4) contains the graviton,
gravitini and a graviphoton (spin 1 field of the gravity multiplet). When
reducing to 4 dimensions, the graviton leads to a graviton plus a vector
and a scalar in 4 dimensions. The gravitino gives an extra spin 1/2, and
the graviphoton gives an extra scalar. Thus, we end up with an extra
vector multiplet. This shows that there is a mapping from a $D=5$
supergravity theory with $n$ vector multiplets to a $D=4$ supergravity
theory with $n+1$ vector multiplets. The same procedure leads further
(after the duality transformations of the vectors) to a $D=3$
supergravity theory with $n+2$ hypermultiplets. For the geometry, this
means a mapping from a real $n$-dimensional manifold, to a complex $n+1$
dimensional manifold, which is called the $\bf r$-map~\cite{deWit:1992nm},
and from there to a quaternionic $n+2$-dimensional manifold, the
so-called $\bf c$-map~\cite{Cecotti:1989qn}. Note that e.g. in the
5-dimensional theory one has $n+1$ vectors when there are $n$ vector
multiplets. These $n+1$ vectors can better be treated together, rather
than separating them in a graviphoton and $n$ vectors of vector
multiplets. In fact, for duality transformations they should be
considered together. The same applies in 4 dimensions. This remark will
be the starting point in section~\ref{ss:scForExtra}.

\section{The family of special geometries}\label{ss:familyspecial}

We can distinguish between theories that appear in rigid supersymmetry,
and those in supergravity. This leads to the overview in the upper part
of table~\ref{tbl:SpecGeom}.
\begin{table}[ht]
\begin{tabular}{|c|ccc|}
\hline
   & $D=5$ vector multiplets & $D=4$ vector multiplets & hypermultiplets \\
\hline
 rigid & affine  & affine &  \\
 (affine) & very special real & special K\"{a}hler & hyperk\"{a}hler  \\ \hline
 local & (projective) & (projective) &   \\
 (projective) & very special real & special K\"{a}hler  & quaternionic-K\"{a}hler \\
\hline \multicolumn{4}{r}{\epsfxsize=13cm
 \epsfbox{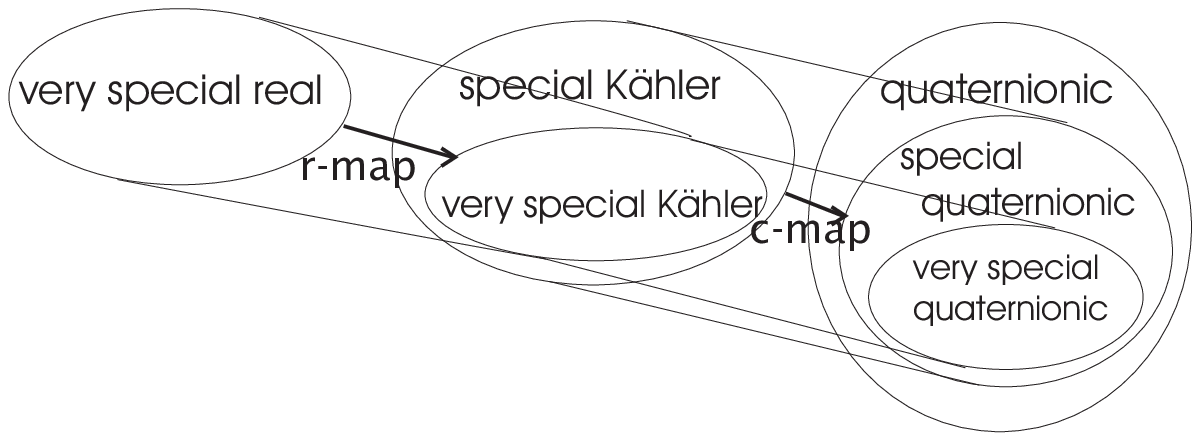} }
\end{tabular}
  \caption{\it Geometries from supersymmetric theories with 8 real
  supercharges, and the connections provided by the ${\bf r}$-map and the
  ${\bf c}$-map.   }\label{tbl:SpecGeom}
\end{table}
The geometries that are related to rigid supersymmetry have been called
`affine' in the mathematics
literature~\cite{Freed:1997dp,Alekseevsky:1999ts}, while those for
supergravity are called `projective' (and these are the default, in the
sense that e.g. special K\"{a}hler refers to the geometry that is found in
supergravity). The analogous manifolds with 3 complex structures got
already a name in the literature: the ones that occur in rigid
supersymmetry are the hyperk\"{a}hler manifolds, while those in supergravity
are the quaternionic-K\"{a}hler manifolds\footnote{Mathematicians include
hyperk\"{a}hler as a special case of what they call `quaternionic-K\"{a}hler',
while physicists reserve the name quaternionic to the manifolds that have
non-vanishing SU$(2)$ curvature, which excludes the hyperk\"{a}hler ones.
Furthermore, we will restrict ourselves to the quaternionic-K\"{a}hler
manifolds of negative scalar curvature, as those are the only ones that
appear in supergravity. For manifolds with continuous isometries, this
implies that they are non-compact.} The manifolds that are defined by the
scalar sector of $N=2$, $D=4$ vector multiplets are called special
K\"{a}hler. When one considers $N=1$ supergravity in 4 dimensions, all K\"{a}hler
manifolds can occur. The presence of 8 real supercharges restricts the
possibilities, and the restricted class is denoted by `special
K\"{a}hler'~\cite{Strominger:1990pd}. Definitions of such manifolds
independent of supergravity were given in~\cite{Craps:1997gp}, and a
review appeared in~\cite{VanProeyen:1999ya}. Another definition has been
given in~\cite{Freed:1997dp,Alekseevsky:1999ts} and has been discussed in
this workshop in the talk of V.~Cort\'{e}s. The real manifolds that appear in
five dimensions got the name of very special real manifolds.
In~\cite{ACCVP} a definition in more mathematical terms is given.

The $\bf r$-map and $\bf c$-map, discussed in the previous section,
induce a terminology for subclasses of special K\"{a}hler and
quaternionic-K\"{a}hler manifolds. The image of the very special real
manifolds under the $\bf r$-map define the very special K\"{a}hler manifolds.
The fact that some of the scalar fields have their origin in components
of vector fields of 5 dimensions leads to isometries for all these
manifolds. Thus, not all special K\"{a}hler manifolds are in this image (as we
will illustrate below for the symmetric spaces), and very special K\"{a}hler
manifolds are a non-trivial subclass of special K\"{a}hler manifolds. The
$\bf c$-map defines in the same way the `special quaternionic-K\"{a}hler'
manifolds as a non-trivial subclass of quaternionic-K\"{a}hler manifolds, and
in an obvious way, also very special quaternionic-K\"{a}hler manifolds are
defined, see the lower part of table~\ref{tbl:SpecGeom}.

Homogeneous and symmetric spaces are the most known manifolds. These are
spaces of the form $G/H$, where $G$ are the isometries and $H$ is its
isotropy subgroup. The group $G$ is not necessarily a semi-simple group,
and thus not all the homogeneous spaces have a clear name. The symmetric
spaces are those for which the algebra splits as $g=h+k$ and all
commutators $[k,k]\subset h$. The homogeneous special manifolds are
classified in~\cite{deWit:1992nm}.

It turns out that homogeneous special manifolds are in one-to-one
correspondence to realizations of real Clifford algebras with signature
$(q+1,1)$ for real, $(q+2, 2)$ for K\"{a}hler, and $(q+3,3)$ for quaternionic
manifolds. Thus, the spaces are identified by giving the number $q$,
which specifies the Clifford algebra, and by specifying its
representation. If $q$ is not a multiple of 4, then there is only one
irreducible representations, and we thus just have to mention the
multiplicity $P$ of this representation. The spaces are denoted as
$L(q,P)$. If $q=4m$ then there are two inequivalent representations,
chiral and antichiral, and the spaces are denoted as $L(q,P,\dot P)$.
\begin{table}[ht]\caption{\it Homogeneous manifolds.
In this table, $q$, $P$, $\dot P$ and $m$ denote positive integers or
zero, and $q\neq 4m$. SG denotes an empty space, which corresponds to
supergravity models without scalars. Furthermore, $L(4m,P,\dot
P)=L(4m,\dot P,P)$. The horizontal lines separate spaces of different
rank. The first non-empty space in each column has rank~1. Going to the
right or down a line increases the rank by~1. The manifolds indicated by a
$\star$ did not get a name.} \label{tbl:homsp}
\begin{center}
\begin{tabular}{|l|ccc|}
\hline
&real & K\"{a}hler & quaternionic \\
\hline &&&\\[-3mm]
$L(-3,P)$&&   & $\frac{\USp(2P+2,2)}{\USp(2P+2)\otimes \SU(2)} $
     \\[2mm]
$SG_4$&&  SG       &$\frac{\U(1,2)}{\U(1)\otimes \U(2)} $
    \\[2mm]
\hline&&&\\[-3mm]
$L(-2,P)$&&$\frac{\U(P+1,1)}{\U(P+1)\otimes \U(1)}$
    &$\frac{\SU(P+2,2)}{\SU(P+2)\otimes \SU(2)\otimes \U(1)} $
        \\[2mm]
$SG_5$&SG  & $\frac{\SU(1,1)}{\U(1)}$
    &$\frac{G_2}{\SU(2)\otimes \SU(2)}$  \\[2mm]
\hline&&&\\[-3mm]
$L(-1,P)$&$\frac{\SO(P+1,1)}{\SO(P+1)}$& $\star$ & $\star$ \\[2mm]
\hline&&&\\[-3mm]
$L(4m,P,\dot P)$& $\star$ & $\star$& $\star$ \\[2mm]
$L(q,P)$&$X(P,q)$&$H(P,q)$&$V(P,q)$\\[2mm]
\hline
\end{tabular}
\end{center}
\end{table}
If we use $n$ as the complex dimension of the special K\"{a}hler space, the
dimension of these manifolds is ($\dot P=0$ if $q\neq 4m$)
\begin{equation}
  n=3+q+(P+\dot P){\cal D}_{q+1}\,,\qquad\left\{
  \begin{array}{l}
    \mathop{\rm dim}\nolimits_\mathbb{R}[\mbox{very special real }L(q,P,\dot P)] =n-1 \\
    \mathop{\rm dim}\nolimits_\mathbb{R}[\mbox{special K\"{a}hler }L(q,P,\dot P)]=2n \\
    \mathop{\rm dim}\nolimits_\mathbb{R}[\mbox{quaternionic-K\"{a}hler }L(q,P,\dot P)]=4(n+1). \
  \end{array}\right.
 \label{nformula}
\end{equation}
where ${\cal D}_{q+1}$ is the dimension of the irreducible representation
of the Clifford algebra in $q+1$ dimensions with positive signature, i.e.
\begin{eqnarray}
&& {\cal D}_{q+1}=1 \ \ \mbox{for }q=-1,0\,,\qquad {\cal D}_{q+1}=2 \ \
\mbox{for }q=1\,,\qquad {\cal D}_{q+1}=4 \ \
\mbox{for }q=2\,,\nonumber\\
&&{\cal D}_{q+1}=8 \ \mbox{for }q=3,4\,,\qquad{\cal D}_{q+1}=16 \
\mbox{for }q=5,6,7,8\,,\qquad {\cal D}_{q+8}=16\, {\cal D}_q\,.
 \label{dimcalDq}
\end{eqnarray}
The very special manifolds are defined by coefficients $C_{IJK}$ as we
will see below. For the homogeneous ones, we can write them as
\begin{equation}
C_{IJK}\,h^Ih^Jh^K = 3\left\{ h^1\,\big(h^2\big)^2 -h^1\,
\big(h^\mu\big)^2 -h^2\,\big(h^i\big)^2 +\gamma_{\mu ij}\,h^\mu\, h^i\,h^j
\right\}.\label{soldhom}
\end{equation}
We decomposed the indices $I=1,\ldots ,n$ into $I= 1, 2, \mu, i$, with
$\mu=1,\ldots, q+1$ and $i=1,\ldots, (P+\dot P){\cal D}_{q+1}$. Here,
$\gamma_{\mu ij}$ is the Clifford algebra representation that we
mentioned. Note that these models have predecessors in 6 dimensions, with
$q+1$ tensor multiplets and $(P+\dot P){\cal D}_{q+1}$ vector multiplets.
The gamma matrices are then the corresponding coupling constants between
the vector and tensor multiplets.

Considering further the table~\ref{tbl:homsp}, we find in the
quaternionic spaces the homogeneous ones that were found
in~\cite{Alekseevsky1975}, together with those that were discovered
in~\cite{deWit:1992nm} (the ones with a $\star$ except for the series
$L(0,P,\dot P)$, which were already in~\cite{Alekseevsky1975}, and
denoted there as $W(P,\dot P)$).

Observe that the classification of homogeneous spaces exhibits that the
quaternionic projective spaces have no predecessor in special geometry,
and that the complex projective spaces have no predecessor in very
special real manifolds. Similarly, only those with $q\geq -1$ can be
obtained from 6 dimensions [with the scalars of tensor multiplets
describing SO$(1,q+1)/$SO$(q)$] and $L(-1,0)$ corresponds to pure
supergravity in 6 dimensions.

There are still symmetric spaces in the range $q\geq -1$. These are shown
in table~\ref{tbl:symvs}.
\begin{table}[ht]\caption{\it Symmetric very special manifolds. Note that
the very special real manifolds $L(-1,P)$ are symmetric, but not their
images under the {\bf r} map.} \label{tbl:symvs}
\begin{center}
\begin{tabular}{|l|ccc|}
\hline
&real & K\"{a}hler & quaternionic \\
\hline
$L(-1,0)$&$\SO(1,1)$&$\left[\frac{\SU(1,1)}{\U(1)}\right]^2$&$\frac{\SO(3,4)}{(
\SU(2)) ^ 3 } $ \\
$L(-1,P)$&$\frac{\SO(P+1,1)}{\SO(P+1)}$& & \\
 $L(0,P)$&$ \SO(1,1)\otimes \frac{\SO(P+1,1)}{\SO(P+1)}$&
$\frac{\SU(1,1)}{\U(1)}\otimes\frac{\SO(P+2,2)}{\SO(P+2)\otimes \SO(2)}$ &
$\frac{\SO(P+4,4)}{\SO(P+4)\otimes \SO(4)} $\\
$L(1,1)$& $\frac{{\mathop{\rm
S}}\ell(3,\mathbb{R})}{\SO(3)}$&$\frac{\Sp(6)}{\U(3)
}$&$\frac{F_4}{\USp(6)\otimes \SU(2)}$\\
$L(2,1)$& $\frac{{\mathop{\rm
S}}\ell(3,C)}{\SU(3)}$&$\frac{\SU(3,3)}{\SU(3)\otimes
\SU(3)\otimes \U(1)}$&$\frac{E_6}{\SU(6)\otimes \SU(2)}$\\
$L(4,1)$& $\frac{\SU^*(6)}{\Sp(3)}$&$\frac{\SO^*(12)}{\SU(6)\otimes
\U(1)}$&$\frac{E_7}{\overline{\SO(12)}\otimes \SU(2)}$\\
$L(8,1)$& $\frac{E_6}{F_4}$&$\frac{E_7}{E_6\otimes
 \U(1)}$&$\frac{E_8}{E_7\otimes \SU(2)}$\\
\hline
\end{tabular}
\end{center}
\end{table}
For the symmetric special K\"{a}hler spaces, this reproduces the
classification obtained in~\cite{Cremmer:1985hc}, while the quaternionic
symmetric spaces contain the `Wolf'-spaces. The full set of isometries
for all the homogeneous manifolds has been obtained
in~\cite{deWit:1993wf}, and all these results have been summarised
in~\cite{deWit:1995tf}.

\section{Superconformal methods for extra symmetry}\label{ss:scForExtra}

We already mentioned that it is advantageous to treat the graviphoton on
equal footing with the other vectors. Indeed, duality transformations can
mix the field strengths of all these vectors. You can thus expect that a
formalism where these are treated democratically will show more of the
symmetry structure. Therefore, we want to use a formalism with $n+1$
vector multiplets (for $n$ denoting the number of physical vector
multiplets). But that would lead to too many spin-1/2 and spin-0 fields.
Let us concentrate first on $D=5$. A formalism with $n+1$ vector
multiplets would thus have also $n+1$ scalars, of which only $n$ can be
physical. One scalar should therefore be a gauge degree of freedom. The
same applies to one of the fermions. The extra symmetries that are used in
this context can naturally be combined in the supergravity group to
enlarge it to a superconformal group.

What we have in mind can be illustrated first for pure gravity. We show
how Poincar\'{e} supergravity is obtained after gauge fixing a conformal
invariant action. The conformal invariant action for a scalar $\phi $ (in
4 dimensions) is
\begin{eqnarray}
 {\cal L} & = & \sqrt{g}\left[  \ft12(\partial _\mu \phi )(\partial ^\mu \phi )-\ft1{12}R\phi
 ^2\right] \,,
 \nonumber\\
 \delta \phi  & = & \Lambda _D\phi \,,\qquad \delta g_{\mu \nu }=-2\Lambda
 _D g_{\mu \nu }\,,\label{confScalar}
 \end{eqnarray}
where the second line gives the local dilatation symmetry that leaves
this action invariant. Now, we can gauge fix this dilatation symmetry by
choosing\footnote{A gauge fixing can be interpreted as choosing better
coordinates such that only one field still transforms under the
corresponding transformations. Then, the invariance is expressed as the
absence of this field from the action. In this case we would use $g'_{\mu
\nu} =g_{\mu \nu }\phi ^2$ as $D$-invariant metric. One can check that
this redefinition also leads to~(\ref{purePoincare}) in terms of the new
field.} the gauge $\phi =\sqrt{6}$. This leads to the pure Poincar\'{e} action
\begin{equation}
  {\cal L}=-\ft1{2}\sqrt{g}R\,.
 \label{purePoincare}
\end{equation}
Pure Poincar\'{e} is in this way obtained from a conformal action of a scalar
after gauge fixing. This scalar, which we will denote further as
`compensating scalar', thus has no physical modes. Observe that the
action~(\ref{confScalar}) describes a scalar with negative kinetic energy
in order that~(\ref{purePoincare}) describes a graviton with positive
kinetic energy. The negative kinetic energy of the compensating scalar
will be important below.

In the full models, this scalar is unified with all the other scalars in
the theory, and this is useful to clarify some isometries of these
theories.

After this motivation for using the superconformal methods, let us repeat
the structure of the superconformal groups that we use. They have 3 parts:
\begin{description}
  \item[conformal group:] consisting of translations and rotations (the Poincar\'{e}
  group), dilatations, and special conformal transformations, i.e. the
  group SO$(D,2)$.
  \item[supersymmetries:] ordinary (that appear also in the super-Poincar\'{e}
  group) and special supersymmetries that are the counterparts of the
  special conformal transformations. For the theories that we consider
  here, there are 8 real components in the ordinary supersymmetries and 8
  in the special ones.
  \item[R-symmetry:] the bosonic group is in general the product of the
  conformal group and an extra bosonic group that rotates the
  supersymmetries. This group appears in the anticommutator of ordinary
  with special supersymmetries. In 5 dimensions this R-symmetry group is
  $\SU(2)$, in 4 dimensions it is $\SU(2)\times \U(1)$ (always for
  theories with 8 ordinary supersymmetries).
\end{description}
The whole supergroup is F${}^2(4)$ (the `2' indicating the real form) for
$D=5$, and $\SU(2,2|2)$ for $D=4$. See other
reviews~\cite{VanProeyen:1999ni,VanProeyen:2001ng} for more details and
references on the supergroups.

\section{Construction in 3 steps}\label{ss:construction3}

We will now review the essential steps in the construction of these
theories. There are 3 steps involved. First, we define multiplets as
representations of the superconformal algebra. Second, we define an
invariant action. This should thus be a superconformal invariant, the
analogue of~(\ref{confScalar}). Finally we gauge-fix the superfluous
symmetries, i.e. all those that belong to the superconformal group and
not to the super-Poincar\'{e} group. These methods are already known
since~\cite{Ferrara:1977ij,Kaku:1978nz,Kaku:1978ea}.

Often these steps are mixed. In many (most) papers one treats at the same
time the action and the transformation laws. We will, however, keep them
strictly separated. This will give insight in what parts of the structure
follows from the algebra, and what follows from the action. We will even
see that in some cases the construction of an action is not possible,
while still we have dynamical equations. In view of this concept, we will
put special emphasis on the hypermultiplets and quaternionic geometry.

\subsection{Multiplets in the superconformal algebra}
We have to warn immediately that the terminology `superconformal algebra'
has to be interpreted with care in field theory. It is not really the
group concept for 2 reasons.

First, the algebra is `soft'. Consider the commutator of two
supersymmetry transformations with parameters $\epsilon _1$ and $\epsilon
_2$ in a gauge theory. The result should be a translation, which on the
scalars is a derivative of the field, but in a gauge theory, derivatives
all appear  as covariant derivative, i.e. we will get
\begin{equation}
 \left[ \delta (\epsilon _1),\delta (\epsilon _2)\right] \phi =\bar \epsilon _1\gamma
^\mu \epsilon _2\,D_\mu \phi\,, \qquad \mbox{with}\qquad D_\mu \phi
=\partial _\mu \phi -A_\mu ^\alpha T_\alpha \phi\,,
 \label{QQcommD}
\end{equation}
where the covariant derivative involves all gauge transformations of the
scalar, $T_\alpha \phi$ multiplied by the corresponding gauge fields
$A_\mu ^\alpha$. The last term thus means that the commutator of two
supersymmetries contains a gauge transformation with parameter $\bar
\epsilon _1\gamma ^\mu \epsilon _2\,A_\mu ^\alpha$. The latter expression
should be the structure `constant'. But it clearly is no constant, rather
it is a `structure function'. Jacobi identities (or their generalizations
in field theory) imply that also other commutators involve fields.
Especially the fields that belong to the multiplet involving the gauge
fields can be expected in structure functions.

A second special feature of the algebra is that it can be `open'. This
terminology is used to indicate that the closure of the algebra involves
equations of motion. In fact, a commutator of two symmetries of an action
should always be a linear combination of symmetries. But there are
trivial symmetries. Any action $S(\phi )$ is invariant under the `trivial'
transformation (here it is presented for bosonic fields, but the
extension to fermionic fields is obvious)
\begin{equation}
  \delta_{\rm triv} \phi^i =\eta ^{ij}\frac{\delta S}{\delta \phi
  ^j}\,,
 \label{deltriv}
\end{equation}
where $\eta ^{ij}$ is any antisymmetric tensor (constant or
field-dependent). Indeed, the transformation of the action is
\begin{equation}
  \delta_{\rm triv} S=\frac{\delta S}{\delta \phi ^i}\,\eta ^{ij}
  \frac{\delta S}{\delta \phi^j}=0\,.
 \label{deltrivS}
\end{equation}
So, the trivial symmetries are part of the full set of symmetries, and in
general a commutator of symmetries only closes modulo trivial
transformations:
\begin{equation}
  \left[ \delta (\epsilon _1),\delta (\epsilon _2)\right] \phi^i
=\mbox{superconformal}+\eta ^{ij}(\epsilon _1,\epsilon _2)\,\frac{\delta
S}{\delta \phi ^j}\,.
 \label{commOpen}
\end{equation}
However, these trivial transformations are clearly model-dependent: they
are determined only when the full action is known. In many cases, we do
not yet want to decide on the full action when we give the transformation
laws. To avoid this situation, one can often enlarge the set of fields by
so-called `auxiliary fields'. With a suitable choice of auxiliary fields
the last term of~(\ref{commOpen}) can be absent, and the transformations
are independent of the action. The algebra closes using only the
symmetries that are the basis of the theory, not using the `trivial' ones.

Such a set of auxiliary fields can not always be found, and is also not
always necessary. When there is no suitable set of auxiliary fields, then
the closure of the algebra already leads to field equations. So, the
transformation laws fix the equations of motion. The transformation laws
can thus only be used with a fixed physical content. When we consider
multiplets that are used in different actions, we do not want to change
their multiplet structure each time. Thus in this case, we need auxiliary
fields, in order to use the multiplet transformations in these different
situations. Fortunately, in the superconformal multiplets that are used
for different actions, we always find a suitable set of auxiliary fields.

After these preliminary remarks on the meaning of `multiplets of the
superconformal algebra' we can discuss the main multiplets. The first one
that we should discuss is the \textit{`Weyl
multiplet'}~\cite{deWit:1980ug}. This is the name that is given to the
multiplet which contains the gauge fields of all the symmetries in the
superconformal algebra. There are 3 types of fields in this multiplet:
\begin{description}
  \item[independent gauge fields:] these are the vielbein $e_\mu ^a$ as
  gauge field of the translations, the gravitini $\psi _\mu ^i$ as gauge
  fields of the ordinary supersymmetries, the field $b_\mu $, gauge field
  of dilatations, and the gauge fields of the R-symmetry $V_\mu ^{(ij)}$.
  \item[dependent gauge fields.] As is well-known already in Poincar\'{e}
  supergravity, the spin-connection, which is the gauge field of Lorentz
  transformations, is not an independent field, but is constrained to be
  some function of the vielbein. In supergravity it is also a function of the
  gravitino, and in the superconformal setup it is a function $\omega _\mu
  {}^{ab}(e_\mu ^a,\psi _\mu ^i,b_\mu )$. Also the gauge fields of special
  supersymmetry and of special conformal transformations are not
  independent fields, but functions of the other fields of the multiplet.
  \item[auxiliary fields] to close the algebra. There are different
  possibilities for the fields that can be used
  here~\cite{Bergshoeff:1986mz,Bergshoeff:2001hc}.
\end{description}
This multiplet satisfies a soft algebra with all the symmetries of the
superconformal algebra. The structure functions between generators of the
algebra thus depend on the fields of this multiplet. For 5 dimensions,
this has been constructed recently
in~\cite{Bergshoeff:2001hc,Fujita:2001kv}. The first reference explains
our methods in more detail and has references to the constructions in
other dimensions.

Then we turn to the \textit{vector multiplets}. As the name says, these
involve vectors. These vectors are gauge fields of symmetries that do not
belong to the superconformal group. They commute with the superconformal
symmetries. The vector multiplets involve also other fields as we saw
before: the `gauginos' and the scalars (and auxiliary fields). The
multiplet is defined `in the background of the Weyl multiplet'. This
means that the transformation laws satisfy the soft algebra that the Weyl
multiplet has defined, but also the gauge transformations of the vectors
appear in the algebra. E.g. (without going into notational details), the
ordinary supersymmetries commute to
\begin{equation}
  \left[ \delta (\epsilon _1),\delta (\epsilon _2)\right] =\,\mbox{superconformal
  algebra with modified }D_\mu
+\delta _G(\bar \epsilon _2\epsilon _1 X)\,,
 \label{QQtoG}
\end{equation}
where the first term represents the algebra as it is satisfied by the
Weyl multiplet, except that covariant derivatives are replaced by an
expression as in~(\ref{QQcommD}) including the new gauge transformations.
These gauge transformations also appear in the last term, where the
parameter contains $X$, the scalar of the multiplet. Observe that, in a
solution of the theory with non-vanishing fields of the vector multiplet,
these extra terms give rise to central charges.

Auxiliary fields can be introduced to close the algebra. Note the
hierarchy in the multiplets. We first had to define the Weyl multiplet
before we could introduce the vector multiplets. This is so because the
Weyl multiplet does not involve the gauge transformations gauged by the
vectors of the vector multiplets. On the other hand, the fields in the
vector multiplets transform of course under the superconformal
symmetries, and we thus need the Weyl multiplet to be able to introduce
the vector multiplets.

Then we can introduce \textit{hypermultiplets}. These are defined in the
background of the Weyl multiplet and possibly also in the background of
the vector multiplet [this is illustrated at the end of this section,
see~(\ref{delQzetagauge}) and~(\ref{delQzetagaugeConf}]. The latter is the
case if one considers hypermultiplets that transform non-trivially under
the gauge transformations of the vector multiplets. Auxiliary fields to
close the algebra (in the sense explained before that `open' means closed
including the trivial symmetries) exist for the simplest quaternionic
manifolds, or can be introduced if one uses the methods of harmonic
superspace. However, we can avoid this. We do not need auxiliary fields
any more at this point. This is because the hypermultiplets are at the
end of the hierarchy line. We are not going to introduce any further
multiplet in the background of the hypermultiplets, as these do not
introduce new gauge symmetries. When we considered the vector multiplets,
the construction had to take into account that the multiplets can be used
for various possible actions (including hypermultiplets or not). When we
come to the hypermultiplets, however, we know what we are looking for. The
hypermultiplet transformations define one particular physical system.

The algebra thus closes only modulo equations of motion. This means that
we have already the physical theory although we have no action yet. The
multiplet and the transformation rules are defined, and as we will
illustrate below, the complex structures are defined on this manifold.
But there is no metric. This corresponds to the notion of a hypercomplex
manifold in the Mathematics literature~\cite{Salamon:1986}. Homogeneous
examples of hypercomplex manifolds that are not hyperk\"{a}hler were
constructed by~\cite{Spindel:1988sr,Joyce:1992,Barberis:1996} and some
constructions of non-homogeneous hypercomplex manifolds were proposed by
D.\ Joyce~\cite{Joyce:1992}, A.\ Swann and H.\
Pedersen~\cite{QuatWorksh2}. Various aspects have been treated in two
workshops with mathematicians and
physicists~\cite{QuatWorksh1,QuatWorksh2}.

As this aspect is new, let us make it more explicit. We start with $4r$
scalar fields and $2r$ spinors. The scalars are denoted by $q^X$ with
$X=1,\ldots ,4r$ and the spinors by $\zeta ^A$ with $A=1,\ldots ,2r$. We
use the formulation in 5 dimensions, and first for rigid supersymmetry.
For details on spinor properties and our conventions, we refer
to~\cite{Bergshoeff:2001hc}. We consider general transformations for the
scalars under the two supersymmetries with parameters $\epsilon ^i$,
$i=1,2$:
\begin{equation}
  \delta_Q (\epsilon) q^X= - \rmi \bar\epsilon^i \zeta^A f_{iA}^X(q)\,,
 \label{delQq}
\end{equation}
where the `vielbeins' $f_{iA}^X(q)$ satisfy a reality condition
\begin{equation}
\rho _A{}^BE_i{}^j  (f^X_{jB})^*=f^X_{iA}\,,
 \label{vpro-realfrho}
\end{equation}
defined by matrices $E_i{}^j$ and $\rho _A{}^B$ that satisfy
\begin{equation}
  E\,E^*=-\unity _2\,, \qquad \rho \,\rho ^*=-\unity _{2r}\,.
 \label{vpro-EErhorho}
\end{equation}
One may choose a standard antisymmetric form for $\rho $ and identify $E$
with $\varepsilon $ by a choice of basis. The transformations on
variables with an $A$ index are by the reality condition restricted to
G$\ell (r,\mathbb{Q})=\,\SU ^*(2r)\times\U(1)$.

The supersymmetry algebra can be closed on the scalars by choosing the
supersymmetry transformations of the fermions as
\begin{equation}
 \delta_Q (\epsilon) \zeta^A = \ft12 \rmi \slashed{\partial} q^X
f_X^{iA}(q)\epsilon_i - \zeta^B \omega_{XB}{}^A(q)\left[  \delta
(\epsilon) q^X \right] ,
 \label{delQzeta}
\end{equation}
where the $f$ that appear here have indices in opposite position as
in~(\ref{delQq}), indicating that they are the inverse matrices as
$4r\times 4r$ matrices
\begin{equation}
   f_{Y}^{iA}  f_{iA}^X=\delta _Y{}^X\,,\qquad f_{X}^{iA}  f_{jB}^X=\delta
   _j{}^i \delta _B{}^A\,.
 \label{vpro-ff1}
\end{equation}
The functions $\omega_{XB}{}^A(q)$ in~(\ref{delQzeta}) appear in an
integrability condition:
\begin{equation}
\mathfrak{D}_Y f_{iB}^X \equiv
\partial_Y f_{iB}^X - \omega_{YB}{}^A(q)   f_{iA}^X + \Gamma _{ZY}^X(q) f_{iB}^Z =
0\,,
 \label{Df0}
\end{equation}
where $\Gamma _{ZY}^X(q)=\Gamma _{YZ}^X(q)$ is any symmetric function.

This can then be interpreted geometrically. $\omega_{XB}{}^A(q)$ are seen
as gauge fields for the G$\ell (r,\mathbb{Q})$. Obviously, we can
interpret $\Gamma _{YZ}^X(q)$ as an affine connection. Also complex
structures can be defined as ($\alpha =1,2,3$ and using the three sigma
matrices)
\begin{equation}
  J_X{}^{Y\alpha }\equiv -\rmi f_X^{iA}(\sigma ^\alpha )_i{}^jf_{jA}^Y\,.
 \label{vpro-defJf}
\end{equation}
The complex structures satisfy, due to (\ref{vpro-ff1}), the quaternion
algebra
\begin{equation}
  J^\alpha J^\beta =-\unity _{4r}\delta ^{\alpha \beta}
  +\varepsilon ^{\alpha \beta\gamma }J^\gamma\,.
 \label{vpro-defJ}
\end{equation}
This defines the space of the scalars to be a hypercomplex manifold.

One can then check that with the given transformations and
identities~(\ref{vpro-ff1}) and~(\ref{Df0}) the supersymmetry algebra on
the fermions closes modulo terms proportional to
\begin{equation}
\Gamma^A \equiv  \slashed{\mathfrak{D}} \zeta^A + \ft12 \varepsilon
^{ij}f^X_{jC} f^Y_{iD} {\cal R}_{XY}{}_B {}^A
 \zeta^B\bar{\zeta}^D \zeta^C\,.\label{eqmozeta}
\end{equation}
There are more terms if the fields transform under the gauge group of a
vector multiplet, see~\cite{Bergshoeff:2001prep}. Putting this equal to
zero (demanding an on-shell algebra) gives an equation of motion for the
fermions. The supersymmetry transformation of this equation gives then
also an equation of motion for the bosons. As announced before, we thus
have already physical equations despite the absence of an action.

All this can be generalized to a superconformal setup. The essential new
ingredient is then a `homothetic Killing vector'
$k^X$~\cite{deWit:1998zg}, which in 5 dimensions satisfies
\begin{equation}
  \mathfrak{D}_Y k^X \equiv \partial_Y k^X + \Gamma _{YZ}^X k^Z =
\ft32 \delta_Y{}^X\,.
 \label{homotheticKilling}
\end{equation}
The presence of this vector allows one to extend the transformations of
rigid supersymmetry to the superconformal
group~\cite{deWit:1998zg,deWit:1999fp,Bergshoeff:2001prep}, with e.g.
transformations under the dilatations and SU$(2)$ R-symmetry group:
\begin{equation}
  \delta _{D,\SU(2)}(\Lambda _D,\Lambda ^\alpha)q^X= \Lambda _D k^X+ \ft23\Lambda ^\alpha
  k^YJ^\alpha{}_Y{}^X\,.
 \label{delDSU2q}
\end{equation}

To illustrate the concept of the background of vector multiplets and Weyl
multiplet for the hypermultiplet, we can give the full forms
of~(\ref{delQzeta}). The given formula is applicable for rigid
supersymmetry. If the hypermultiplet transforms moreover under a gauge
transformation, $\delta q^X=\Lambda ^Ik_X^I(q)$ where $\Lambda ^I$ are
the parameters, and this gauge transformation is associated to the vector
multiplet whose scalars are $h^I$, then the transformation is
\begin{equation}
\delta_Q (\epsilon) \zeta^A = \ft12 \rmi \slashed{D} q^X
f_X^{iA}(q)\epsilon_i - \zeta^B \omega_{XB}{}^A(q) \left[  \delta
(\epsilon) q^X \right] +\ft12 h^I k_I^X f_X^{iA}\epsilon _i\,.
 \label{delQzetagauge}
\end{equation}
The covariant derivative includes now a term $-A_\mu ^Ik_I^X$. If we
consider the multiplet in the local superconformal background, then the
transformation is
\begin{equation}
\delta_Q (\epsilon) \zeta^A = \ft12 \rmi \slashed{D} q^X
f_X^{iA}(q)\epsilon_i - \zeta^B \omega_{XB}{}^A(q) \left[  \delta
(\epsilon) q^X \right]+\ft13\gamma ^{ab}T_{ab} k^Xf_X^{iA}\epsilon
_i+\ft12 h^I k_I^X f_X^{iA}\epsilon _i\,,
 \label{delQzetagaugeConf}
\end{equation}
where $T_{ab}$ is one of the auxiliary fields of the Weyl multiplet. Also
here, the latter term can be present or not depending on whether the gauge
symmetry is gauged or not.

We thus stress that all this is obtained independent of an action. A
similar result can be obtained for tensor multiplets in 5 dimensions. As
mentioned, in the Abelian case they are dual to vector multiplets.
However, in the non-Abelian theory, the inclusion of these multiplets
leads to more general possibilities~\cite{Gunaydin:1999zx}. These
multiplets are also at the end of the hierarchy as that they are not used
for the definition of further multiplets. As for hypermultiplets, their
algebra is not closed. We thus again obtain equations of motion without
the presence of an action~\cite{Bergshoeff:2001prep}. This mechanism of
having physical theories without an action is a concept that is known in
other cases too. E.g. with self-dual tensors, an action is difficult to
construct, and as such, type IIB supergravity~\cite{Schwarz:1983wa} is a
first example.\footnote{There are methods to have an action for these
theories, either by breaking explicit Lorentz invariance, or by including
extra auxiliary fields and gauge
transformation~\cite{Pasti:1997vs,Dall'Agata:1998va}, but these theories
can be, and usually are, considered without a proper action.} Theories
without action are also often used in the group manifold approach, see
e.g.~\cite{D'Auria:1983xj}.

\subsection{Define an invariant action}
To define an invariant action, we need in each case some extra ingredient
apart from the transformation laws. E.g. in 4 dimensions one can start
from a holomorphic function $F(X)$ of the complex scalars of these
multiplets. This is the object that in superspace would be integrated
over the chiral superspace. Thus given the $F(X)$, the action is
determined. For conformal symmetry $F$ should be homogeneous of
degree~2~\cite{deWit:1984pk}.

For vector multiplets in 5 dimensions, we still need a constant tensor
$C_{IJK}$ where $I,J,K$ label the vector multiplets. If we want to use
their duals, the tensor multiplets, then we need also an antisymmetric
metric $\Omega _{MN}$, where $M,N$ label the tensor multiplets. As this
has to be non-degenerate, it implies that the tensor multiplets appear
with an even number (and they are in symplectic representations of the
gauge group). Observe that this requirement only comes from the
requirement of an action, and can thus be avoided if one does not insist
on the existence of an action.

For the hypermultiplets one also needs an antisymmetric metric in the
fermion sector $C_{AB}$. The supersymmetry algebra on the hypermultiplets
closes only modulo functions $\Gamma ^A$ (\ref{eqmozeta}), and we now
want to interpret them as field equations. The relation~(\ref{deltriv})
says that these functions should be proportional to the Euler-Lagrange
equations with antisymmetric coefficients\footnote{It was antisymmetric
for bosons and should be symmetric for fermions, but a charge conjugation
matrix that is implicit in taking the derivative w.r.t. $\bar \zeta $,
which is antisymmetric, implies that the $C_{AB}$ should be
antisymmetric.}
\begin{equation}
  \frac{\delta S}{\delta \bar{\zeta}^A} = 2C_{AB} \Gamma^B\,.
 \label{ELCGamma}
\end{equation}
The consistency equations of field theory (which one can get e.g. from
the field-antifield formalism) imply that $C_{AB}$ are covariantly
constant, and can be chosen to be constant. The requirement that this
tensor is invariant reduces the symmetry group from $\SU(2)\times
\,{\mathop{\rm G}}\ell (r,\mathbb{Q})$ to $\SU(2)\times \USp (2p,2q)$.
Note that the $\SU(2)$ is not gauged in rigid supersymmetry. In the
(local) superconformal approach, there is the $\SU(2)$ gauge field in the
Weyl multiplet, but that is so far independent of the scalars that
constitute the manifold. We thus have here a hyperk\"{a}hler manifold.

The metric is given by
\begin{equation}
   g_{XY}=f_{X}^{iA}f_Y^{jB}\varepsilon _{ij}C_{AB}\,.
 \label{metrichkfromC}
\end{equation}
Its signature is given by the product of $C_{AB}$ with the matrix $\rho
_A{}^B$ that determined the reality properties, see
e.g.~(\ref{vpro-realfrho}).

\subsection{Gauge fixing}

Finally we have to gauge fix the symmetries of the superconformal algebra
that do not belong to the super-Poincar\'{e} algebra. We are most interested
in the bosonic part, and thus have to consider the gauge fixing of
dilatations and of the R-symmetry group.\footnote{Gauge fixing of the
special conformal transformations is done by fixing the value of the
gauge field of dilatations $b_\mu $.} But one extra ingredient is the
field equation of an auxiliary field of the Weyl multiplet, which is
called $D$.

In 5 dimensions for a general theory with vector multiplets and
hypermultiplets, we can fix the dilatations by fixing the value of one of
the scalars of the vector multiplets, as it was already foreseen in
section~\ref{ss:scForExtra}. The scalars of the hypermultiplets transform
under the $\SU(2)$ R-symmetry group, as we saw in~(\ref{delDSU2q}). Thus
the $\SU(2)$ gauge fixes 3 scalars of a hypermultiplet. The field
equation of the auxiliary field $D$ eliminates the fourth. In 4
dimensions, the R-symmetry group has also a $\U(1)$ factor, and the
scalars of the vector multiplet are complex, such that also there one
vector multiplet loses his scalar by the gauge fixing.  Thus the scalars
of one vector multiplet and one hypermultiplet get fixed. The
corresponding spin-1/2 fields are fixed by a similar procedure: the gauge
fixing of the $S$-supersymmetry and the field equation of a fermion
auxiliary field of the Weyl multiplet.

The conclusion is thus that one full hypermultiplet and one vector
multiplet (apart from its vector being the graviphoton) is `compensating'.
We thus have to start from the Weyl multiplet, $n+1$ vector multiplets
and $r+1$ hypermultiplets to end up with the super-Poincar\'{e} gravity
coupled to $n$ vector multiplets and $r$ hypermultiplets. Similar as we
mentioned in the remark after~(\ref{purePoincare}), we have to start with
negative kinetic terms for the compensating vector multiplets to end up
with a super-Poincar\'{e} theory with positive kinetic terms.

In practice, the field equation of the auxiliary field $D$ looks similar
as a gauge condition of dilatations. In fact, one has a linear
combination of this field equation and the dilatational gauge condition
for the vector multiplet and another linear combination for the
hypermultiplet. We may consider these as two independent scaling
conditions.

\begin{figure}[ht]
\unitlength=1mm
\begin{center}
\begin{picture}(80,30)(0,0)
\put(0,0){\leavevmode \epsfxsize=4.5cm
 \epsfbox{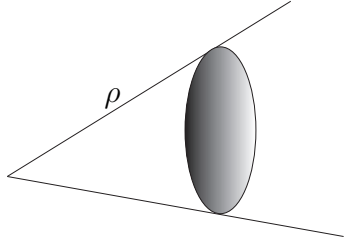}}
\put(13,18){$\rho $}
\end{picture}
\caption{\it Projective very special geometry in the cone of the
$(n+1)$-dimensional space. \label{fig:cone}}
\end{center}
\end{figure}

Let us now see what this means for the geometry. Consider first the $D=5$
vector multiplets. Before the gauge fixing one has a $n+1$ vector
multiplets with scalars $h^I$ and a geometry that is determined by the
$C_{IJK}$ coefficients, i.e. $\rmd s^2=C_{IJK}h^I\rmd h^J\rmd h^k$. This
is an affine geometry. Then the gauge fixing is imposed to fix the scale
transformations of the $h^I$. It is most appropriate to choose as gauge
condition~\cite{Kugo:2000af} a fixed value for the length $\rho \equiv
C_{IJK}h^I h^J h^K$. Notice that this is a dimensionful quantity, and
this gauge fixing thus introduces a length scale that is the gravitational
constant.\footnote{More details on this way of presenting the procedure
are given in~\cite{Kallosh:2000ve}.} The `projective very special
geometry' thus lives on the slice $\rho =M_P^3$, as illustrated in
figure~\ref{fig:cone}.

A similar gauge fixing was performed for vector multiplets in 4
dimensions~\cite{deWit:1984pk}. In this case, the slice as in
figure~\ref{fig:cone} leads to a manifold on which there is a $\U(1)$
structure. This is a Sasakian manifold. Once also the field equation of
the $\U(1)$ gauge field of the Weyl multiplet is used, the geometry is
modified, and the K\"{a}hler curvature is non-zero. The manifold in fact
becomes a K\"{a}hler-Hodge manifold. This is the projective special K\"{a}hler
geometry.

In quaternionic geometry, we start similarly from a hyperk\"{a}hler manifold
with a homothetic Killing vector (a `hyperk\"{a}hler
cone')~\cite{deWit:2001dj}. The slice that is taken here is determined by
a condition as
\begin{equation}
  k^X g_{XY} k^Y=M_P^3\,.
 \label{kkMP}
\end{equation}
This slice has still an $\SU(2)$ symmetry and defines a tri-Sasakian
manifold~\cite{Gibbons:1998xa,deWit:1999fp}. Now the $\SU(2)$ gauge field
of the Weyl multiplet gets by its field equation a value $V_\mu
^\alpha=\omega _X^\alpha(q)\partial _\mu q^X +\ldots $, where $\omega
_X^\alpha(q)$ becomes the gauge field of an $\SU(2)$ in the scalar
manifold. As such, the manifold gets non-trivial $\SU(2)$ curvature and is
promoted to a quaternionic-K\"{a}hler manifold.

With this procedure we can associate the picture in
table~\ref{tbl:constrSpecGeom}, similar to table~\ref{tbl:SpecGeom},
illustrating the mapping from the affine geometries [with signature
$(-+\ldots +)$] to projective geometries.
\begin{table}[ht]
\begin{tabular}{|c|ccc|}
\hline
   & $D=5$ vector multiplets & $D=4$ vector multiplets & hypermultiplets \\
\hline
 conformal & affine  & affine &  \\
 (affine) & very special real & special K\"{a}hler & hyperk\"{a}hler  \\[3mm]
 & $\ \ \ \downarrow$ \parbox{0.7in}{gauge fix\\ $D$} &$\ \ \ \downarrow$ \parbox{0.7in}{gauge fix\\ $D,\,\U(1)$}
 &$\ \ \ \downarrow$ \parbox{0.7in}{gauge fix\\ $D,\,\SU(2)$}\\[4mm]
 Poincar\'{e} & (projective) & (projective) &   \\
 (projective) & very special real & special K\"{a}hler  & quaternionic-K\"{a}hler \\
\hline
\end{tabular}
  \caption{\it Construction of projective special geometries from conformal
  affine geometries, with signature $(-+\ldots +)$.   }\label{tbl:constrSpecGeom}
\end{table}

\section{Conclusions} \label{ss:conclusions}

Special geometry is thus a framework of related theories whose structure
is restricted by 8 real supersymmetry transformations. The restrictions
that are inherent to these models make them particularly useful for
investigations in issues such as dualities. The superconformal
construction gives insight in hidden symmetries that these theories
possess. This construction leads at the end to general actions for
matter-coupled supergravity theories~\cite{deWit:1985px}. We have
schematically indicated here how these can be obtained from the picture in
table~\ref{tbl:constrSpecGeom}. Similar results have been obtained using
other methods. In particular the classic papers for general couplings in
4 and 5 dimensions are~\cite{Andrianopoli:1997cm,AnnaGianguido} used the
group manifold approach. In this paper we did not spend much attention to
the way in which the gauging acts on the hypermultiplets, although this
is automatically obtained by having the hypermultiplets defined `in the
background' of the gauge symmetries of the vector multiplets, as shortly
illustrated by~(\ref{delQzetagauge}). More details on this issue have
been given in~\cite{deWit:2001bk}.

\medskip
\section*{Acknowledgments.}

\noindent This review covers work done with a lot of collaborators, who I
thank for the many interesting discussions and the pleasant atmosphere.
Especially I thank B.~de Wit, with whom most of the present work was
developed. The presentation in this paper has been inspired by several
discussions with D. Alekseevsky, V. Cort\'{e}s and C. Devchand in the context
of the DFG Schwerpunktprogramm (1096) "Stringtheorie im Kontext von
Teilchenphysik, Quantenfeldtheorie, Quantengravitation, Kosmologie und
Mathematik".

\providecommand{\href}[2]{#2}\begingroup\raggedright\endgroup

\end{document}